\documentclass[11pt,nocontents]{iopart}

\usepackage{caption}
\usepackage[latin1]{inputenc}
\usepackage[T1]{fontenc}
\usepackage[english]{babel}
\usepackage{epsfig}
\usepackage{geometry}
\usepackage{bm}
\usepackage{graphicx}
\usepackage{ulem}
\usepackage{inputenc}
\usepackage{amssymb}
\makeindex 

\newcommand{\lambdabar}{{\hbox{$\lambda$\kern-1.ex\raise+0.45ex\hbox{--}}}}

\newcommand{\mnue}{\mbox{$m_{\nu_e}$}}
\newcommand{\mnuezero}{\mbox{$m_{\nu_e,0}$}}
\newcommand{\mnuetwo}{\mbox{$m^2_{\nu_e}$}}
\newcommand{\mnuetwozero}{\mbox{$m^2_{\nu_e,0}$}}

\newenvironment{list2}{
  \begin{list}{$\bullet$}{%
      \setlength{\itemsep}{0in}
      \setlength{\parsep}{0in} \setlength{\parskip}{0in}
      \setlength{\topsep}{0in} \setlength{\partopsep}{0in}
      \setlength{\leftmargin}{0.2in}}}{\end{list}}

\begin{document}

\title{Analysis of KATRIN data using Bayesian inference}

\author{Anna Sejersen Riis}
\address{Department of Physics and Astronomy, University of Aarhus \\
Ny Munkegade, DK-8000 Aarhus C, Denmark}
\address{Institute of Nuclear Physics, Westf{\"a}lische Wilhelms-Universit{\"a}t\\
Wilhelm-Klemm-Str. 9, D-48149 M{\"u}nster}
\author{Steen Hannestad}
\address{Department of Physics and Astronomy, University of Aarhus \\
Ny Munkegade, DK-8000 Aarhus C, Denmark}
\author{Christian Weinheimer}
\address{Institute of Nuclear Physics, Westf{\"a}lische Wilhelms-Universit{\"a}t\\
Wilhelm-Klemm-Str. 9, D-48149 M{\"u}nster}

\ead{\mailto{asr@phys.au.dk},\mailto{sth@phys.au.dk}, \mailto{weinheimer@uni-muenster.de}}

\begin{abstract}

The KATRIN (KArlsruhe TRItium Neutrino) experiment will be analyzing the tritium beta-spectrum to determine the mass of the neutrino with a sensitivity of $0.2$ eV (90\% C.L.). This approach to a measurement of the absolute value of the neutrino mass relies only on the principle of energy conservation and can in some sense be called model-independent as compared to cosmology and neutrino-less double beta decay.

However by model independent we only mean in case of the minimal extension of the standard model. One should therefore also analyse the data for non-standard couplings to e.g. righthanded or sterile neutrinos. As an alternative to the frequentist minimization methods used in the analysis of the earlier experiments in Mainz and Troitsk we have been investigating Markov Chain Monte Carlo (MCMC) methods which are very well suited for probing multi-parameter spaces. We found that implementing the KATRIN $\chi^2$- function in the COSMOMC package - an MCMC code using Bayesian parameter inference - solved the task at hand very nicely.
\end{abstract}
\maketitle
\newpage

\section{Introduction}

The Karlsruhe Tritium Neutrino Experiment KATRIN \cite{Osipowicz:2001sq, KAT:04} will be the first beta decay experiment attempting to measure the electron neutrino mass with sub-eV precision. Presently the experiment is commissioned to start data-taking in 2013/14 and has a projected sensitivity of 0.2 eV (90\% C.L.) to the neutrino mass.

KATRIN is the successor of the experiments in Mainz \cite{Kraus:05} and Troitsk \cite{lobashev03} and will be using some of the same techniques as those. For the technical details of KATRIN see e.g. \cite{otten08}.

Strictly speaking when measuring the 'electron' neutrino mass with $\beta$-decay spectra, what we get is the socalled kinematic neutrino mass. That is, the incoherent sum of neutrino mass eigenvalues weighted by the appropriate entries in the lepton mixing matrix:
\begin{equation*}
 \centering
m^2(\nu_e)=\displaystyle\sum\limits_{i=0}^n |U_{ei}^2| m_i^2.
\end{equation*}

However, because the mass differences between the active neutrino mass states are known to be smaller than KATRINs sensitivity the experiment can effectively only see one mass state (the mass squared differences are $\Delta m_{12}^2=8\times 10^{-5}$ eV$^2$ and $\Delta m_{23}^2=|2.6\times 10^{-3}|$ eV$^2$, respectively \cite{pdg10}). This mass state is sometimes called the 'electron' neutrino mass, but in principle the tritium beta-spectrum could contain the signatures of more than one mass state or of couplings to other particles entirely. In order to be called truly model independent KATRIN's final data should be analyzed also for alternative scenarios - beyond the minimal extension of the standard model.

Performing an analysis for non-standard couplings to the electron neutrino adds more parameter space to the $\chi^2$-function of the experiment. One should therefore consider how an extended analysis should be performed on the KATRIN output in order to get reliable results.

We present here one approach which seems to give several advantages over the standard frequentist analysis. In section 2 we describe our analysis methods before presenting results for a number of cases in section 3. Finally we give some concluding remarks in the last section.

\section{Methods: Frequentist and Bayesian analysis tools}
\subsection{The principles of Mainz and KATRIN data-analysis}

Let us begin by summarising the procedures for production and analysis of KATRIN spectra as performed by a toy model Monte Carlo and analysis code for KATRIN-like experiments \cite{KAT:04}. This code has previously been used to forecast the experiment's sensitivity to the neutrino mass~\cite{Masood:07}.

Because KATRIN has an integrating spectrometer (a consequence of the MAC-E (Magnetic Adiabatic Collimation with Electrostatic) filter technique) the beta-spectrum must be written as an integral over the electron energy:
\begin{equation}
\centering
N_s(qU,E_0,m^2_{\nu_e})=N_{tot}\cdot t_U\int_0^{E_0}\frac{dN_{\beta}}{dE_e}(E_0,m^2_{\nu_e}) \cdot f_{res}(E_e,qU)dE_e
\label{eq:tid}
\end{equation}

Here $U$ is the retarding potential of the spectrometer, $N_{tot}$ is the total number of tritium nuclei in the source, $t_U$ is the measurement time allotted for a given value of the retarding potential and $f_{res}$ is the experimental response function (which in turn is a combination of the electron energy loss function of the tritium source and the transmission function of the spectrometer).$\;\frac{dN_{\beta}}{dE_e}$ is the theoretical beta spectrum rate folded with the electronic final state distribution of molecular tritium.

A retarding voltage-independent background rate of $B_0$ is now added to Eq.
(\ref{eq:tid}):
\begin{equation}
	N_b=B_0 \cdot t_U \, .
\end{equation}

This gives us the following theoretical expression for a KATRIN-like spectrum:
\begin{equation}
\centering
N_{th}(qU,E_0,m^2_{\nu_e}) = N_s(qU,E_0,\mnuetwo ) + N_b \, .
\label{eq:best1}
\end{equation}

Individual spectra (to resemble the real measurements) are built using initial parameters \mnuetwozero\ and $E_{0,0}$ for the neutrino mass squared endpoint energy, respectively. To the theoretical expression is then added a random component from a gaussian distribution with $\sigma(qU_i) =\sqrt{N_s+N_b}$ and $\mu(qU_i)=N_s+N_b$:
\begin{equation}
\centering
N_{exp}(qU)= N_s(qU,E_{0,0},\mnuetwozero) +  N_b + \mathrm{Rnd}\left(Gauss(\sigma,\mu) \right).
\label{eq:best2}
\end{equation}

When we want to fit our randomized beta-spectra we have to account for statistical
fluctuations by allowing the overall amplitude $A$ of the signal as well as the
background rate $B$  to vary against the theoretical
amplitude $A_0$ and background rate $B_0$. In addition, we allow the neutrino
mass squared \mnuetwo\ as well as the endpoint energy $E_0$ to deviate from the
initial parameters of the simulation \mnuetwozero\ and $E_{0,0}$.
\begin{equation}
\centering
N_{fit}(qU,A, B, E_0,m^2_{\nu_e}) =
   A \cdot \frac{N_s(qU,E_0 , m^2_{\nu_e})}{A_0}
   	+ B \cdot \frac{N_b}{B_0} \, .
\label{eq:best3}
\end{equation}

Combining Eq.'s ~(\ref{eq:best2}) and ~(\ref{eq:best3}) we finally get KATRIN's $\chi^2$-function \cite{KAT:04}:
\begin{equation}
\centering
\chi^2(A,B,E_0,\mnuetwo)= \sum_i \left( \frac{N_{exp}(qU_i) -N_{fit}(qU_i,A,B,E_0,\mnuetwo )}{\sigma(qU_i)}\right)^2 \, .
\end{equation}

\begin{figure}[htb!]
\includegraphics[width=15.8cm]{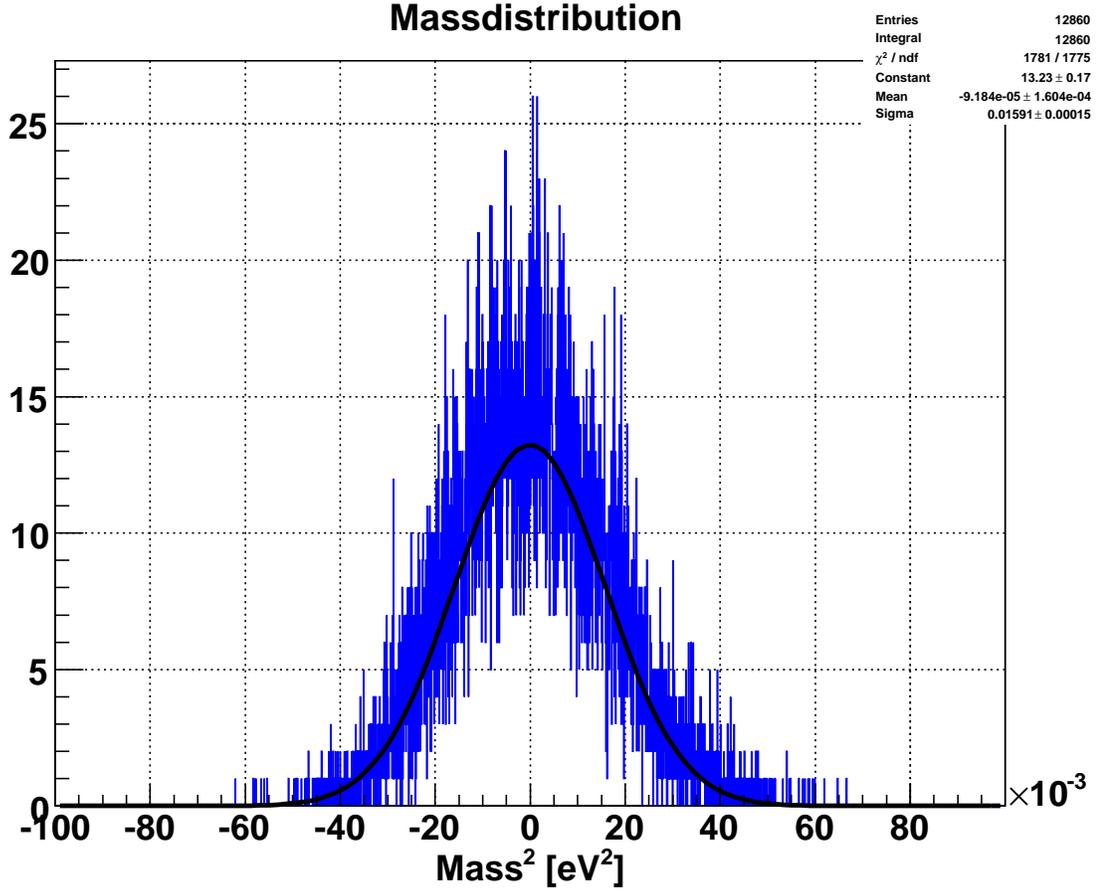}
\caption{An example histogram depicting the neutrino mass-squared values from minimizations of $12860$ beta-spectra\label{fig:fig1} for a KATRIN-like parameter set and an
assumed value for the neutrino mass of $m_{\nu_e} = 0$ measured with an optimized time
distribution over the last 25~eV of the beta-spectrum, e.g. compare to \cite{Masood:07}.}
\end{figure}
The analysis of the simulated data can be performed with Minuit2 which is imbedded in the ROOT-package. This procedure performs a minimization of the $\chi^2$-function using CombinedMinimizer. CombinedMinimizer in turn uses either an evaluation of the covariance matrix or a simplex method to find the best minimum of the $\chi^2$-function in the parameter space \cite{ROOT}. One can now do a standard frequentist analysis to find the statistical uncertainty on e.g. the neutrino mass by producing a suitable amount of Monte Carlo spectra, performing the minimization for each of them and finally inspecting the resulting histograms. An example is shown in Figure ~\ref{fig:fig1} for 12860 spectra produced with $m_{\nu} =$ 0.0 eV.

However as previously indicated the minimization approach has a number of drawbacks. For one thing it does not give any information on multiple minima, and it is not well suited for finding shallow minima. Furthermore extracting detailed information on correlated parameters is pretty laborious. Still the method works just fine for the four well-known free parameters used in a standard KATRIN analysis - see Table ~\ref{tab:org}. But as one adds more parameters the minimization procedure often becomes problematic and rather slow.

\subsection{Bayesian parameter inference with COSMOMC}

As an alternative approach we have considered Markov Chain Monte Carlo and bayesian inference techniques as in the publicly available COSMOMC analysis package for cosmology. Typical cosmological models contain $\sim$ 8-12 parameters and COSMOMC is well suited for relatively fast analysis of such multiparameter spaces \cite{COSMOMC}. The programme is built for analysis of large cosmological datasets such as CMB data from WMAP and supernova surveys but can in principle analyse whatever dataset the user provides - the cosmology can be 'turned off' if it is irrelevant.

COSMOMC uses bayesian statistics for the analysis. When doing socalled bayesian parameter inference one is interested in knowing the posterior probability, $P(\overline{\theta}|D,M)$ - the probability of the \textbf{parameters}, $\overline{\theta}$, given the data $D$ and the model $M$. The 'inverse' question is for the probability of the \textbf{data}, D, given the parameters and the model, $P(D|\overline{\theta},M)$ - this is simply the likelihood function. With these two probabilities and the well-known Bayes theorem,
\begin{equation}
\centering
P(A \wedge B)=P(A)\cdot P(B|A)=P(B)\cdot P(A|B),
\end{equation}

one can write an expression for the posterior probability:
\begin{equation}
\centering
P(\overline{\theta}|D,M) =\frac{L(D|\overline{\theta},M)\cdot \pi(\overline{\theta}|M)}{\varepsilon(D|M)}.
\end{equation}

Here $L$ is the likelihood - which can be easily derived from the $\chi^2$-function\footnote{The
likelihood function $L$ is connected to the $\chi^2$-function as in the following way: $L  = \exp(-\chi^2/2)$.}. The posterior probability is thus proportional to the likelihood.

Meanwhile $\pi(\overline{\theta}|M)$ is the socalled prior probability sometimes referred to as the subjective input - it is what we believe we know from theory before even taking the data into account. Correspondingly this probability has no dependence on the data. Note that we have been using flat priors on all input parameters in this paper.

Finally $\varepsilon(D|M)$, the evidence, is in effect only a parameter-independent normalization constant \footnote{However, it can be important in some contexts, e.g.\ in comparing two qualitatively different models}\cite{Hamann:08}.

When we want to know best-fit values and confidence levels of specific parameters we can then simply integrate over all the remaining (nuisance) parameters. This is called marginalization and the output is called the marginalized probability for the parameter of interest.

In addition to this rather convenient production of parameter probability distributions, from the bayesian inference approach, COSMOMC gives us another great advantage by using a Markov Chain Monte Carlo (MCMC) to probe the parameter space. This will provide a very thorough and easy to inspect mapping of the parameter space of interest.

The purpose of the MCMC is to probe the whole parameter space in a randomized manner. To achieve this one implements the Metropolis Hastings algorithm \cite{Christensen:2001} consisting of three main steps:

\begin{list2}
\vspace*{.1in}
\item Firstly an initial point, $\overline{\theta_0}$ is chosen.
\item Secondly a step is proposed in some random direction, after which the new point is evaluated: $P(\overline{\theta_i} +\overline{\theta_p})$. Here $\overline{\theta_i}$ means the iterative point, and $\overline{\theta_p}$ is the proposed addition taken from some proposal density.
\item Finally the procedure decides whether or not to take the step. The point $\overline{\theta_i} +\overline{\theta_p}$ is accepted if the posterior probability is improved. That is if
\begin{equation}
\centering
\frac{P(\overline{\theta_i} +\overline{\theta_p})}{P(\overline{\theta_i})} \geq 1.
\end{equation}
If the expression above is  $\leq \, 1$ the step is accepted with some probability $r$ (rejected with probability $1-r$). In this manner we generate a set of points $\{ \overline{\theta_i} \}$, also called a Markov Chain. For the number of points, $N$, going to infinity we thus have a representation of the posterior probability.
\end{list2}
\vspace*{.1in}

The decision procedure of the Metropolis Hastings algorithm allows the chain to wander away from any local minima and thus potentially discover other minima (to a degree determined by the value of r\footnote{In our case r is defined as $e^{\Delta\chi^2/2T}$, with temperature, $T=1$.}. On the other hand it also guarantees that the parameter space near the minima is very well probed. Furthermore one can perform the analysis on a combination of multiple chains - all started at random positions - and get an even better picture of the behavior of the different parameters in the allowed intervals. To get rid of un-physical effects from the random starting points one normally allows for a burn-in -- i.e. the first part of the Markov Chain is removed. In our case the burn-in is 50\% of the sample size.

Before running the programme one must carefully choose stepsizes and parameter ranges. Several settings in both the COSMOMC programme as such and in the parameter files can be tweaked to fit ones purpose.

Unfortunately it is in principle not possible to determine in any absolute terms whether or not a specific chain has converged \cite{Christensen:2001}, but various convergence diagnostics have been developed. For instance when analyzing multiple chains a convergence parameter, $R$, defined as the variance of the chain means divided by the mean of the chain variances, can be evaluated. If $1-R$ is less than some chosen small number (in our case 0.03) this information is interpreted as good convergence.

When COSMOMC has generated the chains we need, the data analysis is performed giving us best-fit values and standard deviations for all the parameters.

Additionally COSMOMC produces a number of useful Matlab-files which can be used to produce $1D$ and $2D$ plots of the marginalized distributions. Inspecting this graphical output allows us to determine if the chains have really converged, whether there are multiple minima and perhaps most importantly it shows parameter correlations right away\footnote{In fact COSMOMC -- or rather GetDist -- produces a multitude of diagnostics files in addition to the graphical output. More information on these can be found on the COSMOMC homepage \cite{COSMOMC}. In addition, it is fairly easy to edit the Getdist programme to produce output files that fits ones purpose}.

If we go through all of this for say a single Monte Carlo generated beta spectrum we get all the nice advantages mentioned above. But the analysis of that one spectrum would take many hours as compared to minutes or seconds with the Minuit2 procedure and we would mostly just have achieved a much slower evaluation of the best fit values for that particular spectrum. However if we in stead use the theoretical beta-spectrum (which should represent the average of infinitely many measurements or Monte Carlo realizations) as our input data - but with Monte Carlo generated errorbars - our best fit values and standard deviations from COSMOMC should correspond to the results of the frequentist approach of building histograms for a very large (going to infinity) number of measurements.

To recap we implemented our $\chi^2$-function for KATRIN-like experiments in COSMOMC and simply turned off cosmology. As data-set we have used the theoretical spectra - for any given model - with the Monte Carlo generated errorbars of the original code. The results will be discussed in the following section.\\

\section{Results}
\subsection{The minimal model}
As a first test of our methods we have attempted to reproduce the KATRIN sensitivity. We thus generated a tritium beta-spectrum using as input so far only four parameters: The electron neutrino mass squared \mnuetwozero , the endpoint of the beta-spectrum $E_{0,0}$\footnote{To suppress the number of
digits, we rather plot and write its deviation from 18575~eV - i.e. $\Delta E_0$}, the background count rate $B_0$
and the signal count rate near the beta-spectrum endpoint, $A_0$ - see eq. (\ref{eq:best1}).

The input signal count rate can be calculated as a combination of the column density of the source and the magnetic fields and cross sections of the spectrometer and source. In our case the count rate near the endpoint $E_0$ is included in the analysis code via an amplitude factor $A_0$
as in eq. (\ref{eq:best3}). The exact definition and full calculation of this factor $A_0$ is included in Appendix A of reference \cite{Anna}. Given KATRIN's experimental settings the amplitude has the value $A_0 = 477.5$~Hz.

We would like to remark, that the value of the endpoint energy $E_0$ needs to be treated
as a free parameter to produce realistic fits with respect to fitting of \mnuetwo. Up to now the
$^3$He - $^3$H mass difference is known from precision Penning trap experiments with 1.2~eV precision \cite{nagy06},
but already the fits of the experiments at Mainz \cite{Kraus:05} and Troitsk \cite{lobashev03}
would have needed a much more precise input value to justify keeping $E_0$ fixed in the fit.

\begin{table}
\centering
  \begin{tabular}[htb!]{ |l | c | c | r | }
    \hline
    Parameter & unit & typical input value\\
    \hline \hline
    $m_{\nu_e,0}^2$  & eV$^2$ & 0.0 - 1.0  \\ \hline
    $\Delta E_{0,0}$ & eV     & 0.0        \\ \hline
    $B_0$   & Hz     & 0.01       \\ \hline
    $A_0$    & Hz     & 477.5      \\
    \hline
    \end{tabular}
    \caption{KATRIN standard analysis parameters. Please note, that we show and
      plot the deviation of the value of the endpoint of the beta-spectrum
      from the theoretically expected value: $E_0 = 18575$~eV.
  \label{tab:org}}
  \end{table}
\begin{figure}[htb!]
\centering
\includegraphics[scale=0.7]{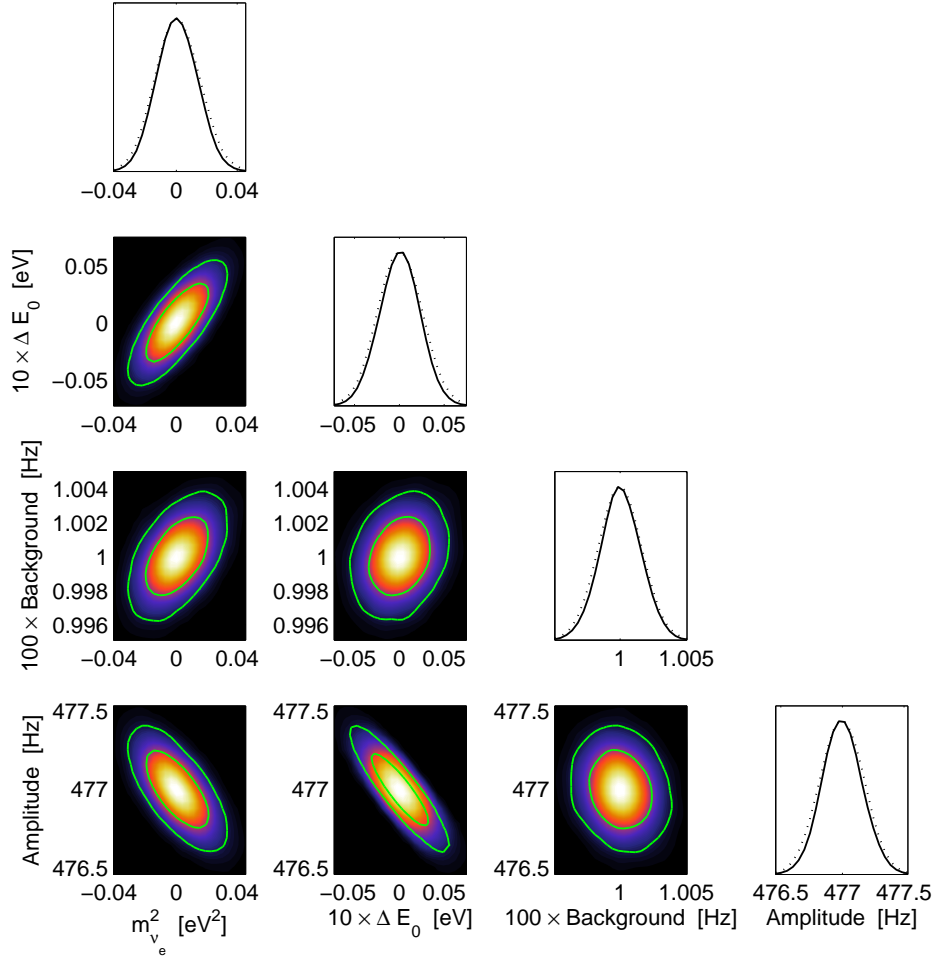}
\caption{A one-neutrino analysis with input mass $\mnuezero=0.0$~eV. In the COSMOMC output the contours (dotted lines) mark the likelihood function in the 2D (1D) -distributions. Color (full lines) mark the data-point distributions.
The results have converged nicely and gives $m_{\nu}^2 =-0.41\cdot 10^{-5}\pm 0.013$ eV.  \label{fig:fig2}}
\end{figure}
The values of our free parameters and the are listed in Table ~\ref{tab:org}.

The COSMOMC results for a theoretical spectrum with $\mnuezero=0.0$ eV are presented in Figure \ref{fig:fig2}. Clearly the chains have converged nicely in this case and the parameters seem well-constrained. The output values of our analysis are

\begin{eqnarray*}
m_{\nu_e}^2   & = & (-0.41\cdot 10^{-5} \pm  0.013) \; {\rm eV} ^2 \\
\Delta E_0  & = & (0.87\cdot 10^{-5} \pm  0.22 \cdot 10^{-2}) \; {\rm eV}  \\
B & = & (1.00\cdot 10^{-2} \pm 0.15 \cdot 10^{-4}) \; {\rm Hz}  \\
A   & = & (477.0 \pm 0.16)  \; {\rm Hz} , \\
 \end{eqnarray*}

We note that our analysis gives a statistical error on the neutrino mass squared of 0.013 eV$^2$, but the statistical uncertainty from the frequentist analysis shown in Figure \ref{fig:fig1} is 0.016 eV$^2$ and thus $\approx$ 23\% larger than our bayesian result \footnote{One can calculate KATRINs sensitivity to the electron neutrino mass at 90\% C.L. using the following equation (and assuming gausssianity):
\begin{equation}
\centering
L=\sqrt{1.64 \sqrt{\sigma_{tot}^2(m_{\nu}^2)}},
\end{equation}
Here $\sigma_{tot}^2(m_{\nu}^2)=\sigma_{stat}^2 + \sigma_{sys}^2$, and KATRIN's systematic error on the neutrino mass squared is the 0.017 eV$^2$ quoted in \cite{KAT:04}. The bayesian analysis gives $\sigma_{stat}^2=$ 0.013 eV$^2$ and a sensitivity of $0.19$ eV (90\% C.L.) on the electron neutrino mass.}.

As a test we therefore calculated the statistical uncertainty for a 11 different values of the neutrino mass in the frequentist approach. A comparison with bayesian results are presented in Table \ref{tab:bayleaf}. We see that we get systematically higher values of the statistical uncertainty when using frequentist methods for the calculation.

\begin{figure}[htb!]
\centering
\includegraphics[width=12.0cm]{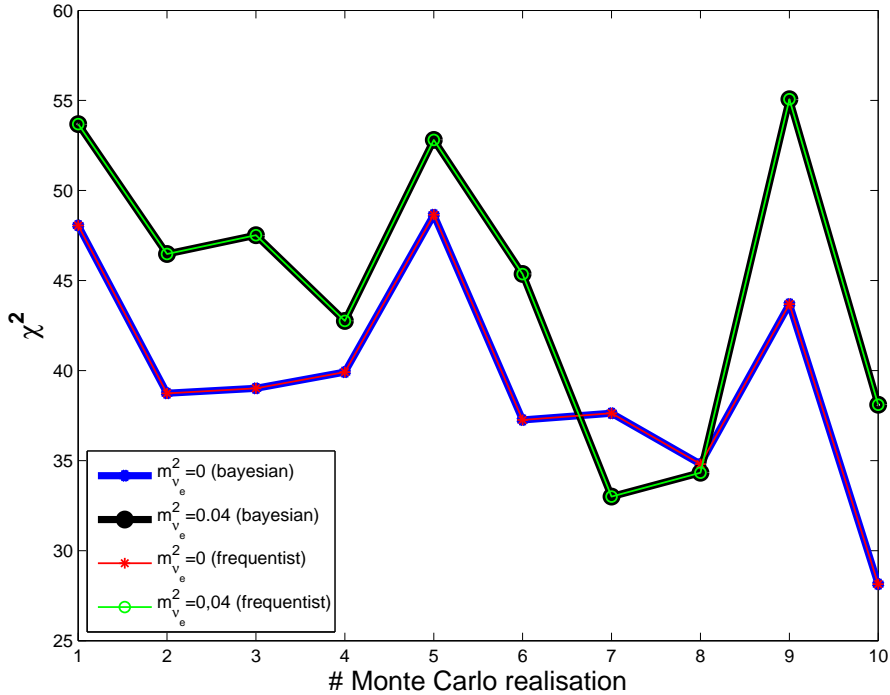}
\caption{The $\chi^2$ value as calculated in COSMOMC (using with flat priors for all input parameters) and ROOT for ten different Monte Carlo realizations of KATRIN-like spectra (36 d.o.f). The results coincide both when the analysis is performed with the mass square fixed at the true input value ($m_{\nu_e}^2=0$ eV$^2$) and when the mass square is fixed at KATRINs sensitivity squared ($m_{\nu_e}^2=0.04$ eV$^2$).  \label{fig:chis}}
\end{figure}

If we try instead to compare the chi-squared values of the two methods the results are more encouraging. We performed the analysis on 10 different Monte Carlo generated spectra keeping the mass square fixed at 0 eV$^2$ (the true input value) and 0.04 eV$^2$ (the sensitivity squared) respectively. This gave identical chi-squared values as shown in Figure \ref{fig:chis}. The relative deviations between the methods were no larger than $\sim 2\cdot 10^{-4}$ for $m_{\nu_e}^2$ fixed at 0 eV$^2$ and $\sim 0.5\cdot 10^{-4}$ for the $m_{\nu_e}^2$ fixed at 0.04 eV$^2$

We are therefore led to believe that the difference in $\sigma_{stat}(m_{\nu}^2)$ is simply down to the use of bayesian instead of frequentist statistics. The procedure with which we find the standard deviations of the parameters (through a marginalization over nuisance parameters) is very different from a frequentist approach. Perhaps it is not too surprising that we do not produce a curve with the exact same specifications as the histogram of Figure \ref{fig:fig1}. We do however get the same $\chi^2$-value as well as the correct outputs and certainly the right trend in the behavior of $\sigma_{stat}(m_{\nu}^2)$. Furthermore we note that our results make no use of any gaussianity assumptions.
In conclusion this bayesian approach seems to produce robust results.\\
\begin{table}
\centering
  \begin{tabular}[htb!]{ |l | c | c | c | c | c | c | c | c | c | c | r | }
    \hline
    $m_{\nu_e}$ [eV] & 0.0 & 0.1 & 0.2 & 0.3 & 0.4 & 0.5 & 0.6 & 0.7 & 0.8 & 0.9 & 1.0 \\ \hline
    $\sigma_{stat, bay}$ [$10^{-2}$ eV$^2$]  & 1.31 & 1.34 & 1.38 & 1.39 & 1.45 & 1.51 & 1.53 & 1.59 & 1.61 & 1.66 & 1.69  \\ \hline
    $\sigma_{stat, fre}$ [$10^{-2}$ eV$^2$]  & 1.64 & 1.70 & 1.79 & 1.95 & 1.88 & 1.91 & 1.93 & 2.04 & 2.02 & 2.09 & 2.21  \\ \hline
    \end{tabular}
    \caption{The statistical uncertainty on the neutrino mass squared for 11 different input values. The second row has been calculated using the bayesian approach in the COSMOMC analysis, while the third row has been calculated in the usual frequentist approach assuming a gaussian distribution function.
  \label{tab:bayleaf}}
  \end{table}

\subsection{Sensitivity plot}

As a first application of our bayesian analysis formalism we have built a sensitivity plot for KATRIN-like experiments. Or rather an illustration of the behavior of the statistical uncertainty on the neutrino mass squared as a function of key experimental settings.

The sensitivity to the electron neutrino mass of a KATRIN-like experiment depends on the signal strength, the background count rate and the energy resolution of the experiment. Given the fairly well understood effect of the background on the sensitivity we investigate only the effect of the the signal strength and the energy resolution - the specific values used are listed in Tables \ref{tab:amps} and \ref{tab:ress}.

Additionally we include an optimization of the measurement time distribution for each of our KATRIN-like experiments (as specified by their amplitude and energy resolution). This optimization has in fact a great deal of influence on the reachable sensitivity of such an experiment. Currently KATRIN is projected to have a runtime of three years but because of experimental stability issues the measurements are performed as a relatively fast scan over the electron energies of interest (or rather retarding voltages) of total duration 966 s. The measurement time allotted to each data point, the $t_U$ in eq. \ref{eq:tid}, has been carefully optimized for KATRIN's experimental settings as described in \cite{KAT:04}.

The basic structure of the measurement time distribution can be represented as three segments around the region of the beta spectrum endpoint.

\begin{list2}
\vspace*{.1in}
\item Firstly one needs measurements up to about 10 eV above the endpoint of the beta-spectrum $E_0$ to determine the correct background.

\item Secondly the 'bulk' region of interest below the endpoint of the beta-spectrum $E_0$ must be treated carefully. Effectively this is the section we optimize.

\item Thirdly, previous investigations by the KATRIN collaboration have pointed out that there exists a region of maximal sensitivity to the neutrino mass. This region is centered around the
electron energy $E_e$ for which the signal counts equals 2 times the background count rates: $N_s=2 N_b$ \cite{Otten:94}. Above this narrow region the total count rate is dominated by background noise. Below it the sensitivity to the neutrino mass drops as one goes away from the endpoint. Therefore extra measurement time is devoted to measurements in this interval. Given the projected KATRIN background count rate of 0.01 Hz the total count rate of the critical point must be 0.03 Hz.
\end{list2}
\vspace*{.1in}

\begin{figure}[htb!]
\centering
\includegraphics[width=8.0cm]{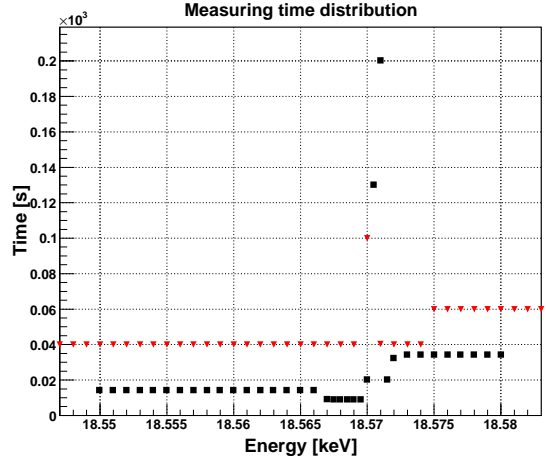}
\caption{The black squares represent the standard KATRIN measurement time distribution and the red triangles show an example of a simplified time distribution with measurement points separated by 1 eV in the main block. This particular time distribution contains only one point in the range: 0.02 Hz < $N_{th}(qU,E_0,\mnuetwo)$ < 0.04 Hz.  \label{fig:time}}
\end{figure}

As a rule we construct the background-block of our distribution as 10 points separated by 1 eV each having $t_U=60$ s. The main block contains a total of 30 points with varied spacing (to be optimized) and $t_U=40$ s. Finally 60 s are added to $t_U$ for 0.02 Hz $\leq \; N_{th}(U) \; \leq$ 0.04 Hz. We find that the 40-100-60 block-structure does a good job of simulating both the structure and total duration of the measurement time distribution while still being simple enough to manipulate. For the sake of comparison we show the fully optimized KATRIN measurement time distribution and one of our simplified distributions in Figure \ref{fig:time}.
\begin{table}
\centering
  \begin{tabular}[htb!]{ |l | c | c | c | c | c | c | c | c | c | r | }
    \hline
    Amplitude  [Hz] & 2.0 & 4.0 & 8.0 & 16.0 & 32.0 & 64.0 & 125.0 & 250.0 & 500.0 & 1000.0 \\ \hline
    \end{tabular}
    \caption{Amplitudes investigated in the sensitivity plot presented in Figure \ref{fig:sensitivity}.
  \label{tab:amps}}
  \end{table}

\begin{table}
\centering
  \begin{tabular}[htb!]{ |l | c | c | c | r | }
    \hline
    Energy resolution  [eV] & 0.5 & 1.0 & 2.0 & 4.0 \\ \hline
    \end{tabular}
    \caption{Energy resolutions investigated in the sensitivity plot presented in Figure \ref{fig:sensitivity}.
  \label{tab:ress}}
  \end{table}

\begin{table}
\centering
  \begin{tabular}[htb!]{ |l | c | c | c | c | c | c | c | c | c | r | }
    \hline
    Data point intervals  [eV] & 0.1 & 0.2 & 0.4 & 0.8 & 1.0 & 1.2 & 1.4 & 1.6 & 1.8 & 2.0 \\ \hline
    \end{tabular}
    \caption{The allowed data point intervals used in the main block of our measurement time distributions.
  \label{tab:ints}}
  \end{table}

To build the sensitivity plot we let a script evaluate $\sigma_{stat}(m_{\nu_e}^2)$ for the combination of all energy resolutions given a specific amplitude. The evaluation include an optimization of the measurement time distribution: For the first energy resolution we find the best statistical uncertainty using all the available distributions (as specified by their data point intervals in the main block - see Table \ref{tab:ints}). We then impose a time saving condition stating that the next energy resolution is allowed only to use the previous best distribution and its closest neighbors. And so on for the remaining energy resolutions. The reason for doing this 'spline' between energy resolutions and not between amplitudes is the expectation that the largest sensitivity fluctuations will take place in the amplitude direction.

\begin{figure}[htb!]
\centering
\includegraphics[width=15.0cm]{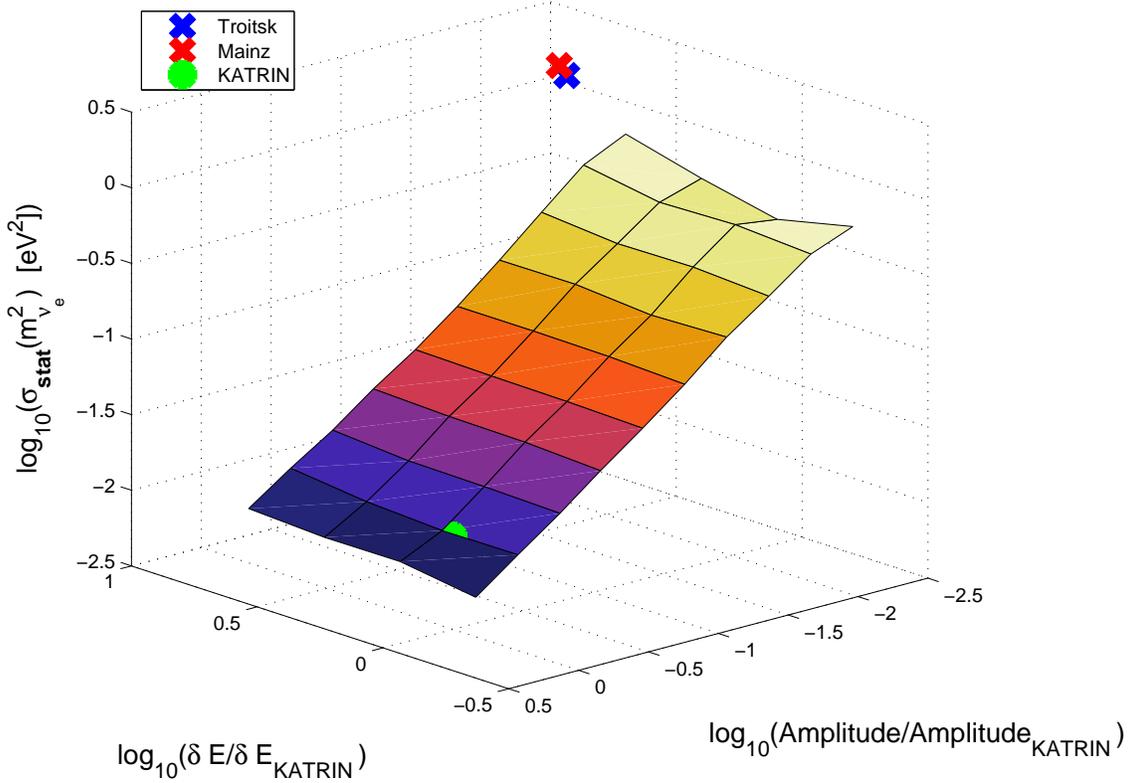}
\caption{Our final sensitivity plot result showing the statistical uncertainty on the neutrino mass squared as a function of amplitude and energy resolution for a KATRIN-like experiment. It is very clear from the figure that $\sigma_{stat}(m_{\nu_e}^2)$ is strongly dependent on the signal count rate (as could be expected) and to a much lesser degree on the energy resolution, $\delta E$. For comparison the figure also show the corresponding statistical uncertainties of the Troitsk, Mainz and KATRIN experiments (The KATRIN-point was calculated with the COSMOMC approach presented here).  \label{fig:sensitivity}}
\end{figure}

We present the average of 10 sensitivity plots as our result in Figure \ref{fig:sensitivity} - including  points corresponding to the statistical uncertainty of the Mainz \cite{Kraus:05} and Troitsk \cite{lobashev03} experiments. The Figure shows a clear log-log dependence of $\sigma_{stat}(m_{\nu_e}^2)$ on the amplitude. The dependence on the energy resolution on the other hand is very weak. Making a fit to the plane of the sensitivity plot gives us:
\begin{equation*}
\log_{10}(\sigma_{stat}(m_{\nu_e}^2))=0.058\cdot \log_{10}(\delta E) - 0.70\cdot\log_{10}(\mathrm{Amp})+0.0038.
\end{equation*}

That is roughly a factor 12 stronger dependency on the amplitude than on the energy resolution in this fit (with $R^2_{fit}$=0.9989). We should note, that these simulations are
done without considering the corresponding systematics, which would most likely give a stronger dependence on the energy resolution  $\Delta E$.

Further it can be noted that the statistical uncertainties from the Troitsk and Mainz experiments (respectively 2.5 and 2.2 eV$^2$ - taken from their final results) are somewhat above the plane. This is as could be expected given the fact that the plot was built specifically from a KATRIN toy model. And as mentioned above we get systematically lower statistical uncertainties using our bayesian analysis algorithm than from corresponding frequentist methods.

Finally we noted a tendency in our procedure to choose large data point separations for the 'optimal' measurement time distribution. This effect - a lower statistical uncertainty caused by analyzing a larger energy interval below the Q-value - was already demonstrated in Figure 19 of \cite{Osipowicz:2001sq}. For our purpose we have kept the systematic uncertainty on the 0.017 eV$^2$ as projected for the KATRIN experiment \cite{KAT:04} and our sensitivity so to speak depends only on the statistical uncertainty.


\subsection{Sterile neutrinos}
After these initial tests of our COSMOMC extension we have tried adding further parameters in our routine.

As previously mentioned it is obvious that KATRIN can not resolve the mass squared differences between the known active states - of $\Delta m_{12}^2=8\times 10^{-5}$ eV$^2$ and $\Delta m_{23}^2=|2.6\times 10^{-3}|$ eV$^2$, respectively. However sterile neutrinos with mass states in the eV range could in principle mix with $\bar\nu_e$. Such neutrinos would provide a much better target for direct detection in beta decay experiments than the active neutrinos which are expected to have sub-eV masses. Their relatively high mass would allow for an easy separation from the primary decay signal in experiments such as KATRIN.

Recently the MiniBooNE collaboration confirmed their previous findings and more indications of a fourth mass state can be found in the so-called reactor anomaly \cite{AAAA:2010,Mention:2011}. Even cosmology suggests an effective number of neutrino species slightly larger than three \cite{Hamann:2010bk,Hamann:2007pi,Hamann:2010pw,GonzalezGarcia:2010un}. We have therefore performed a more thorough investigation of the possible detection potential for sterile neutrinos by KATRIN-like experiments. For the total results see \cite{Anna} and references therein.

Here we will present only our main results for effectively a 1+1 (active + sterile) neutrino scenario. We can justify having only one active neutrino by the known mass squared differences. We kept the active neutrino massless and performed the analysis for a broad range of sterile neutrino masses. For KATRINs standard settings (see Table \ref{tab:org}) we get the result presented in Figure \ref{fig:figa}:

\begin{figure*}[h!]
\centering
\includegraphics[scale=0.45]{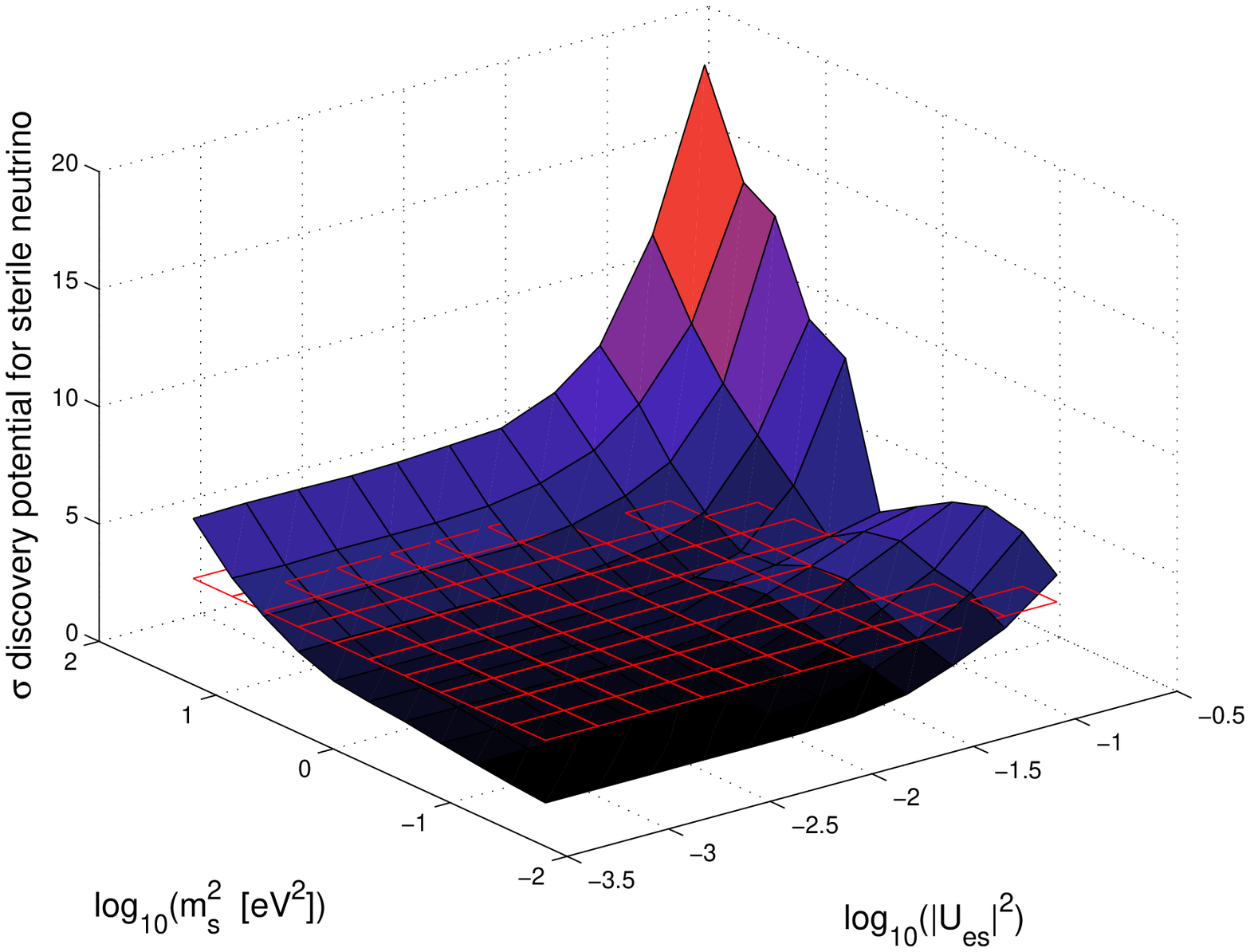}
\hspace{0.01cm}
\includegraphics[scale=0.45]{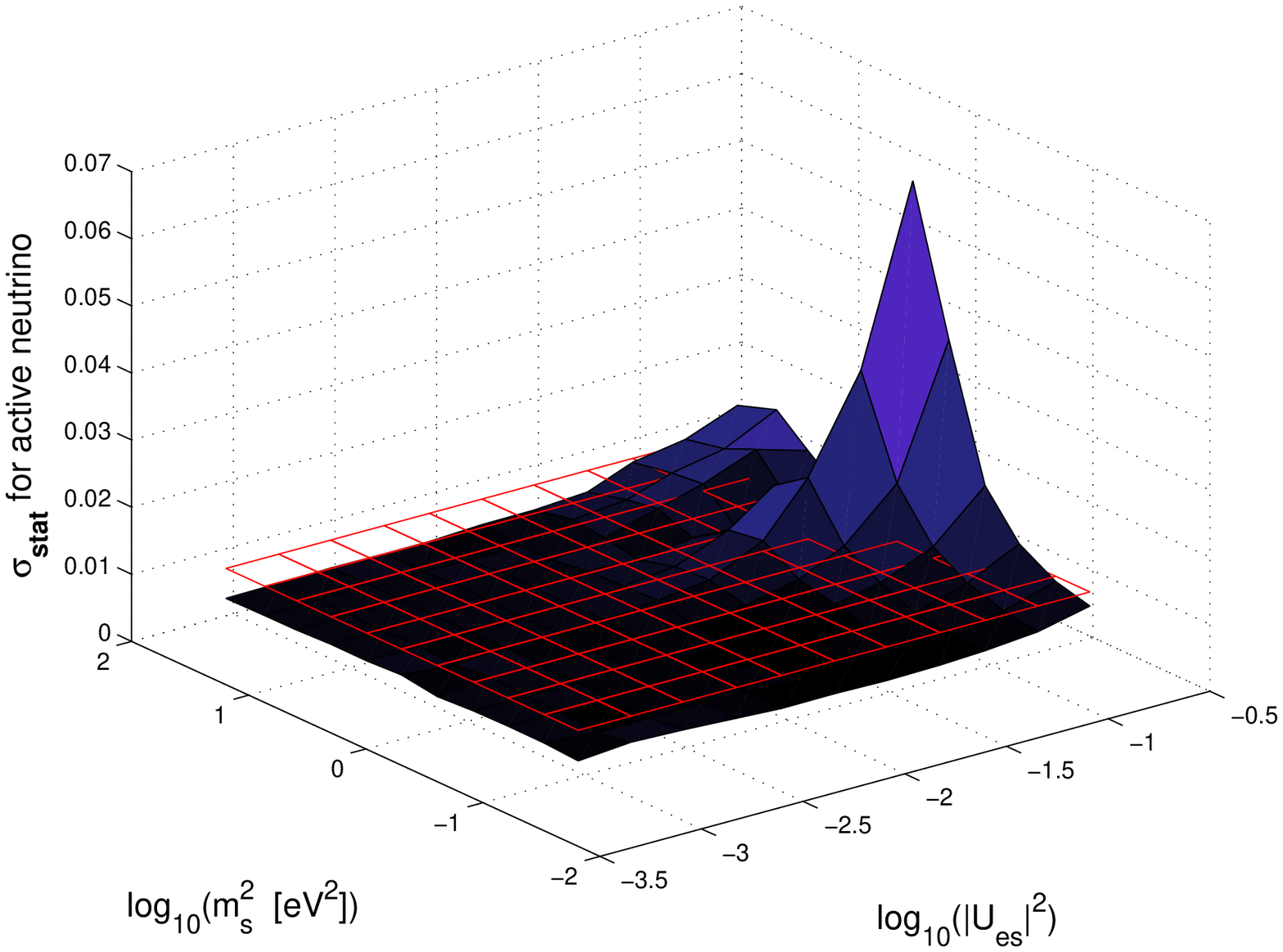}
\caption{The upper figure shows the sigma detection potential of the massive sterile neutrino for the standard KATRIN-like settings, while the lower shows the corresponding statistical deviation (in eV$^2$) on the massless neutrino. The x- and y-axis depicts the logarithm of the sterile mass squared and the mixing weight. The red mesh illustrates the 3$\sigma$ level and the standard one-neutrino statistical uncertainty of around 0.013 eV$^2$ respectively (for this analysis method). As one would expect the mass and the mixing weight must be rather high in order to get a good detection of the sterile component.
\label{fig:figa}}
\end{figure*}

We found that KATRIN should be able to perform a 3$\sigma$ detection of any of the heavy mass states we used as long as $|U_{es}|^2 \gtrsim$ 0.055. Likewise, Correspondingly a 1$\sigma$ detection is achievable for $|U_{es}|^2 \gtrsim$ 0.018.

\subsection{Right handed currents}

As a final application we take a look at righthanded currents. We use the notation of Stephenson {\it{et al}} in \cite{Stephenson:2000mw} and let $b$ parametrise the strength of the right-handed interaction. We define $b = \rho_R \cos \theta_R/\cos \theta$, where $\cos \theta$ is the mixing angle from the mass eigenstate to the left-handed current weak eigenstate and $\cos \theta_R$ is the mixing angle from the mass eigenstate to the corresponding right-handed current weak eigenstate. Again we assume for simplicity that the mass of the electron neutrino can be described by one effective mass eigenstate. $\rho_R$ is the ratio of the effective strength of interactions mediated by righthanded currents to the strength of interactions mediated by the well-known lefthanded weak current. Then the differential beta spectrum is modified in the following way in the presence of righthanded currents \cite{Stephenson:2000mw}\footnote{Note that the last term is equivalent to 2 times the similar expression in \cite{Bonn:2007su} in the limit of small $b$. We wish however to use Eq. \ref{eq:right} retaining the physical meaning of b, rather than treating the coupling as an effective parameter}:
\begin{equation}
\frac{dN_{\beta}}{dE_e}= E_{\nu}\sqrt{(E_{\nu}^2 -m_{\nu}^2)}\left(1 + 2 \frac{b}{1+b^2}\frac{m_{\nu}}{E_{\nu}}\right).
\label{eq:right}
\end{equation}

\begin{figure}[htb!]
\includegraphics[width=15.0cm]{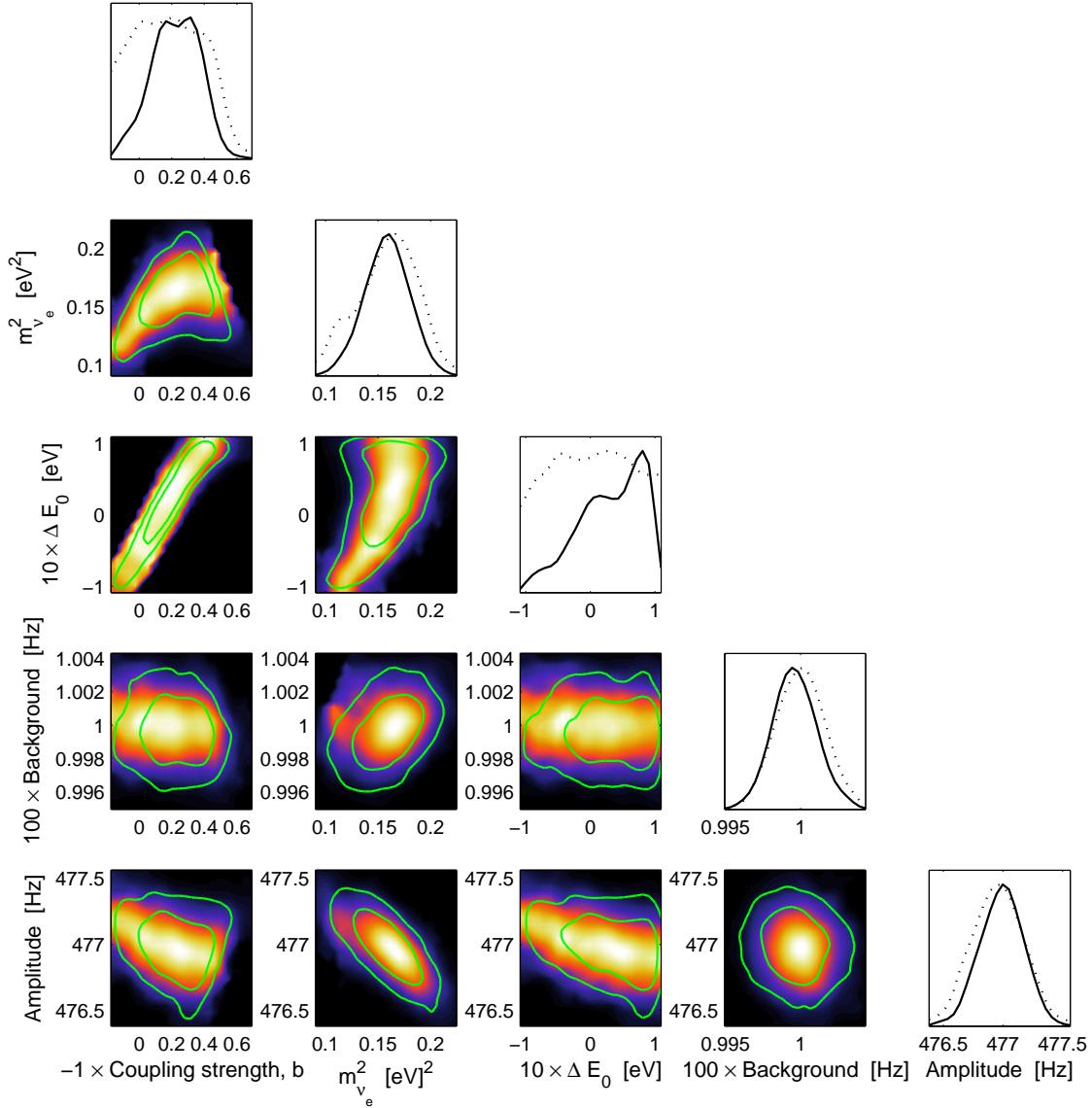}
\caption{A typical COSMOMC output for a model with $b_0=-0.13$ and $\mnuezero=0.4$ eV. One clearly sees the correlations between $m_{\nu_e}$, $b$ and $\Delta E_0$. Despite quite large input parameter ranges for $m_{\nu_e}^2$ and $b$ neither is well constrained.
 \label{fig:f53}}
\end{figure}

From many data of weak precision experiments reviewed in reference \cite{severijns06}
Bonn {\it{et al}} derive an upper limit on $|\frac{2b}{1+b^2}|$ of $0.31$ ($99.7\%$ C.L.)
\cite{Bonn:2007su} translating to $|b| \lesssim 0.16$ in our parameterization \footnote{However it is clear that if $|\frac{2b}{1+b^2}|$ < $0.31$ can also give the limit $|b| \gtrsim 6.3$ and obviously the effect of $b \sim 0$ and $|b| \gg 1$ would give very similar spectra using Eq. \ref{eq:right}. However $b>1$ has no physical meaning cf. our definition}. It is clear from Eq. \ref{eq:right} that the mass and the coupling parameter, $b$, are strongly correlated. Combined with the well known correlation between the neutrino mass squared \mnuetwo\ and the endpoint energy $\Delta E_0$ \cite{otten08} this will propagate to an additional $b$~--~$\Delta E_0$ correlation. Figure \ref{fig:f53} shows a COSMOMC output for all the parameters in a model with input values $b_0=-0.13$ and $\mnuezero=0.4$. The input parameter ranges were $m_{\nu_e, 0}^2 \pm 1.5$ eV$^2$ and $-2.5 < b_0 < 2.5$, respectively. We see that even within these rather large conservative intervals, which were investigated,  neither $m_{\nu_e}$  or $b$ can be determined well.

\begin{figure}[htb!]
\includegraphics[width=7.0cm]{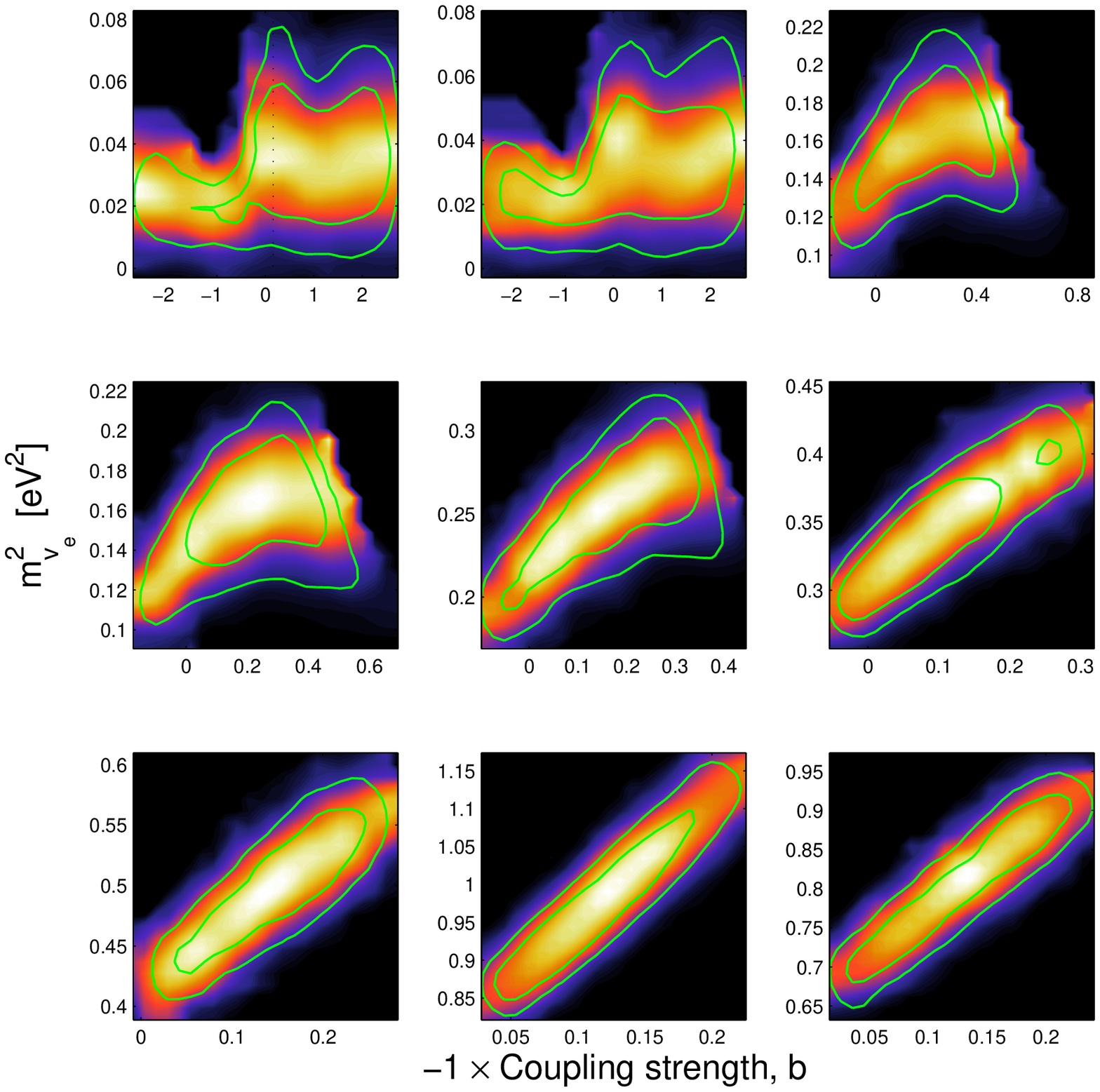}
\hfill
\includegraphics[width=7.0cm]{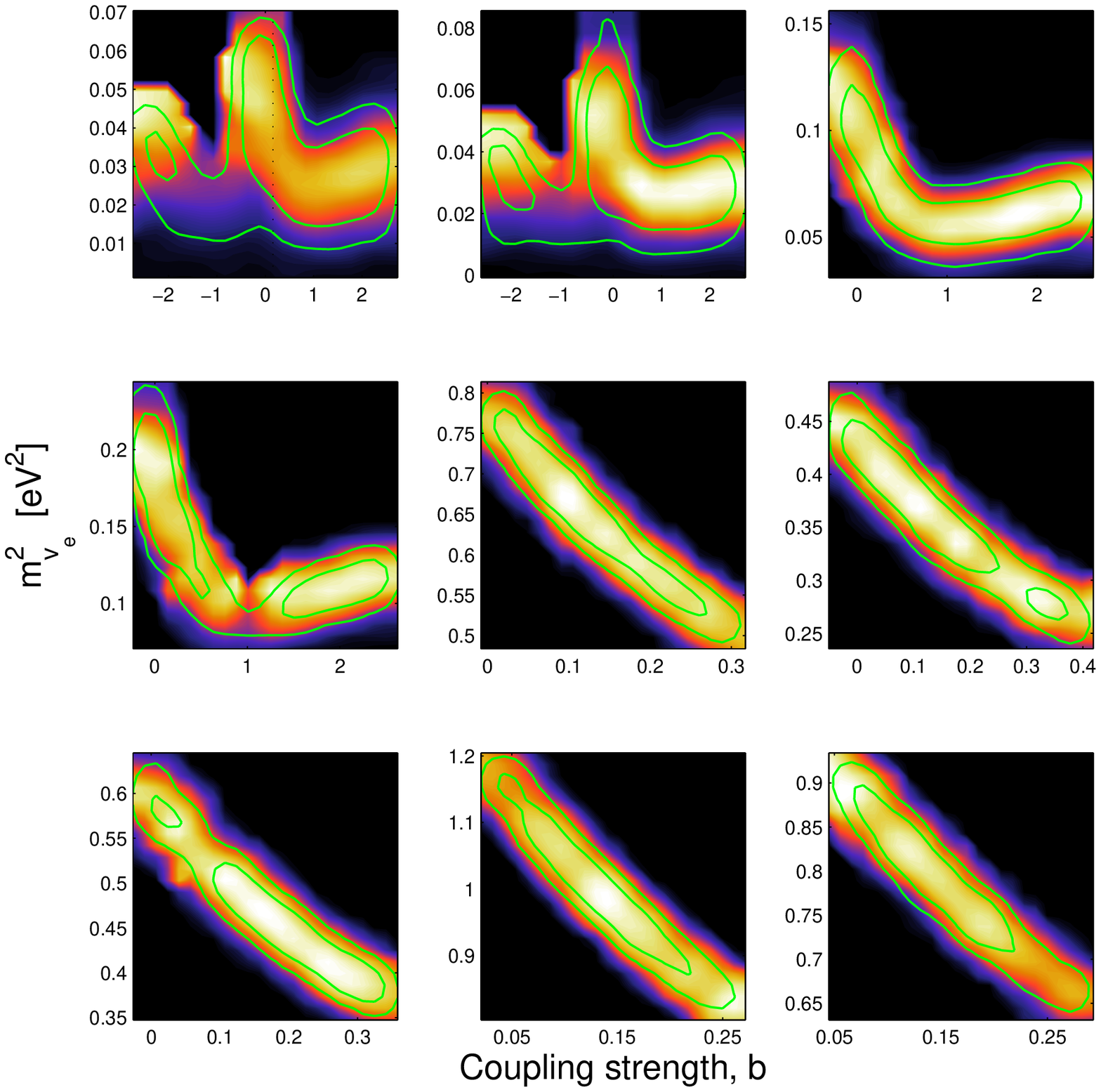}
\caption{The left figure shows marginalized COSMOMC 2D likelihood contours and MCMC data for models with $b_0=-0.13$ and \mnuezero\ ranging from $0.1$ eV to $0.9$ eV (going from the upper left corner to the lower right corner). The right figure has $b_0=0.13$ and the same range of input masses. The correlation between $m_{\nu_e}$ and $b$ only establishes itself when the neutrino mass is larger than KATRINs sensitivity of 0.2 eV. This allows for smaller $\sigma_{stat}(m_{\nu_e}^2)$ than in the case of $b_0=0$ because the MCMC-routine lets the uncertainty on the mass fill the extra parameter dimension ($b$).
 \label{fig:comb3}}
\end{figure}

As another example Figure \ref{fig:comb3} shows the behavior of the \mnue~--~$b$ correlation for a range of masses and $b_0=\pm0.13$. Funnily enough for small input masses this allows for a better $\sigma_{stat}(m_{\nu_e}^2)$ than in the purely lefthanded case,
if the right-handed coupling constant $b$ would be known. We believe this is caused by the large available phase-space in the $b$-direction. This means the likelihood can be made very narrow around the input-value, while still providing many valid solutions. However we also notice that when KATRIN's sensitivity is reached the uncertainty on the mass will be determined again by the experimental limitations and not by numerical solutions. Therefore the correlation between \mnuetwo\ and $b$ reappears.


Given these disconcerting initial results we investigated how large an influence the presence of righthanded currents might have on the output neutrino mass. We have performed the full COSMOMC analysis for the parameters given in Table \ref{tab:Right}. For $b=0$ the analysis was the standard analysis, i. e. without a $b$-dimension. Our main results are presented as a relative bias compared to the $b=0$ -case (with the exclusion of $m_{\nu_e}=0$ in the mass bias, to avoid infinities):
\begin{equation}
\centering
\mathrm{B_{ias}}=100\cdot\frac{X_b-X_{b=0}}{X_{b=0}}
\end{equation}

\begin{figure}[htb!]
\includegraphics[width=7.0cm]{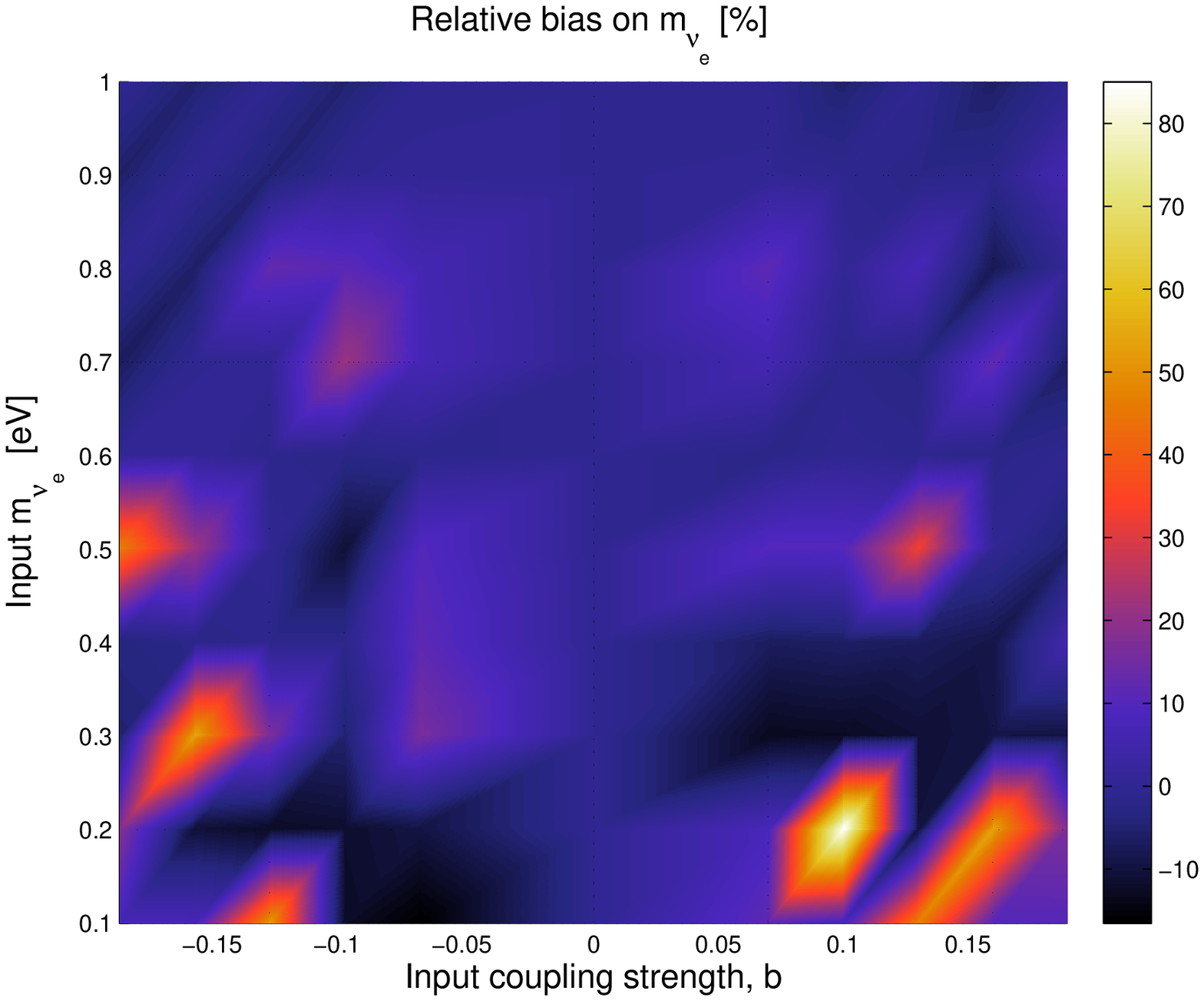}
\hfill
\includegraphics[width=7.0cm]{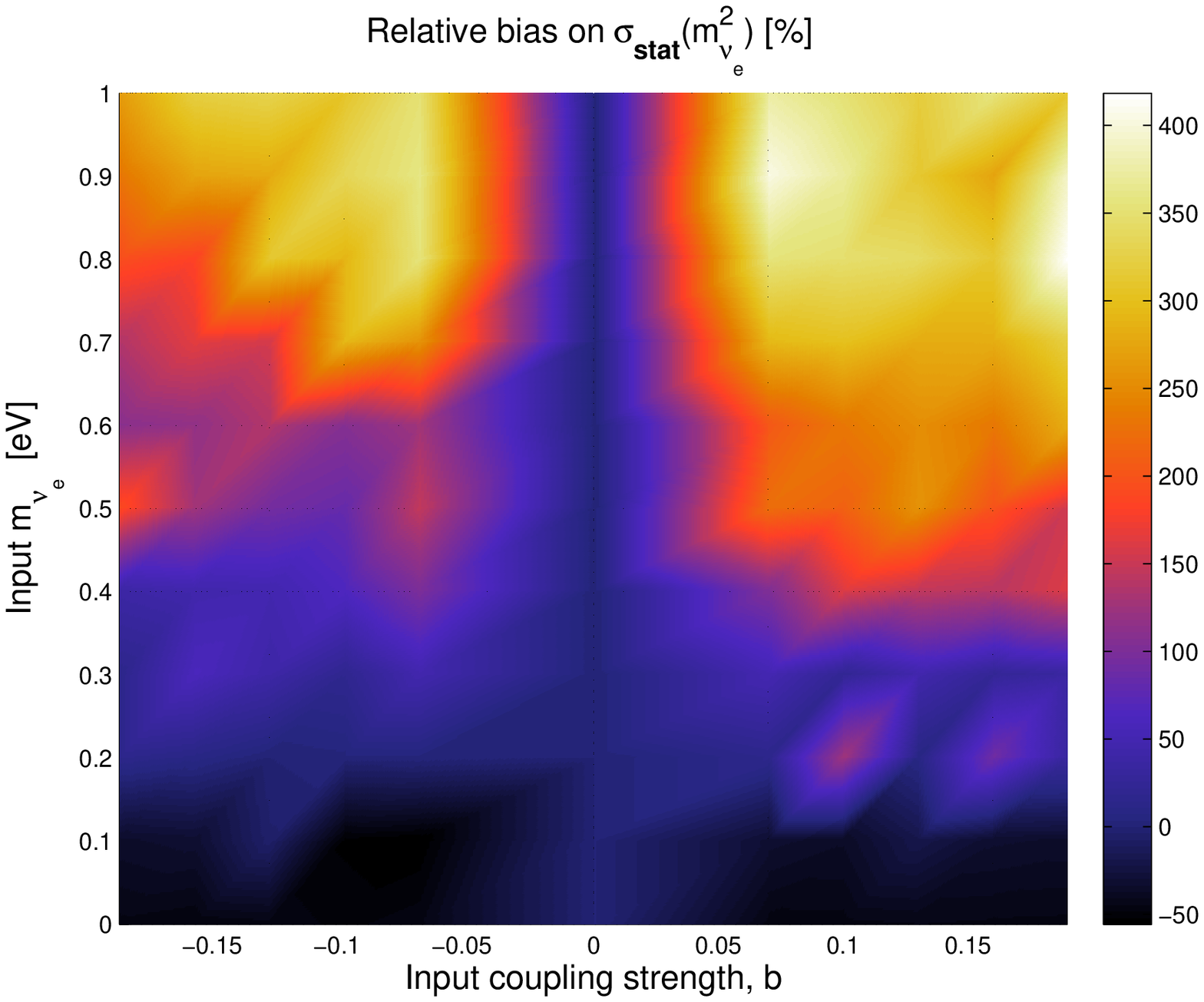}
\caption{The left figure shows the bias on $m_{\nu_e}$ as compared to the case without righthanded couplings and the right figure shows the bias on $\sigma_{stat}(m_{\nu_e}^2)$. This analysis includes the righthanded coupling strength as a free parameter. The bias on the mass is as large as 80\% while the values of $\sigma_{stat}(m_{\nu_e}^2)$ is up to five times as large as for the standard case (barring the parameter range below KATRINs sensitivity where the uncertainty on the mass parameter migrates into the b-dimension in the MCMC).
 \label{fig:m_r}}
\end{figure}

\begin{figure}[htb!]
\includegraphics[width=7.0cm]{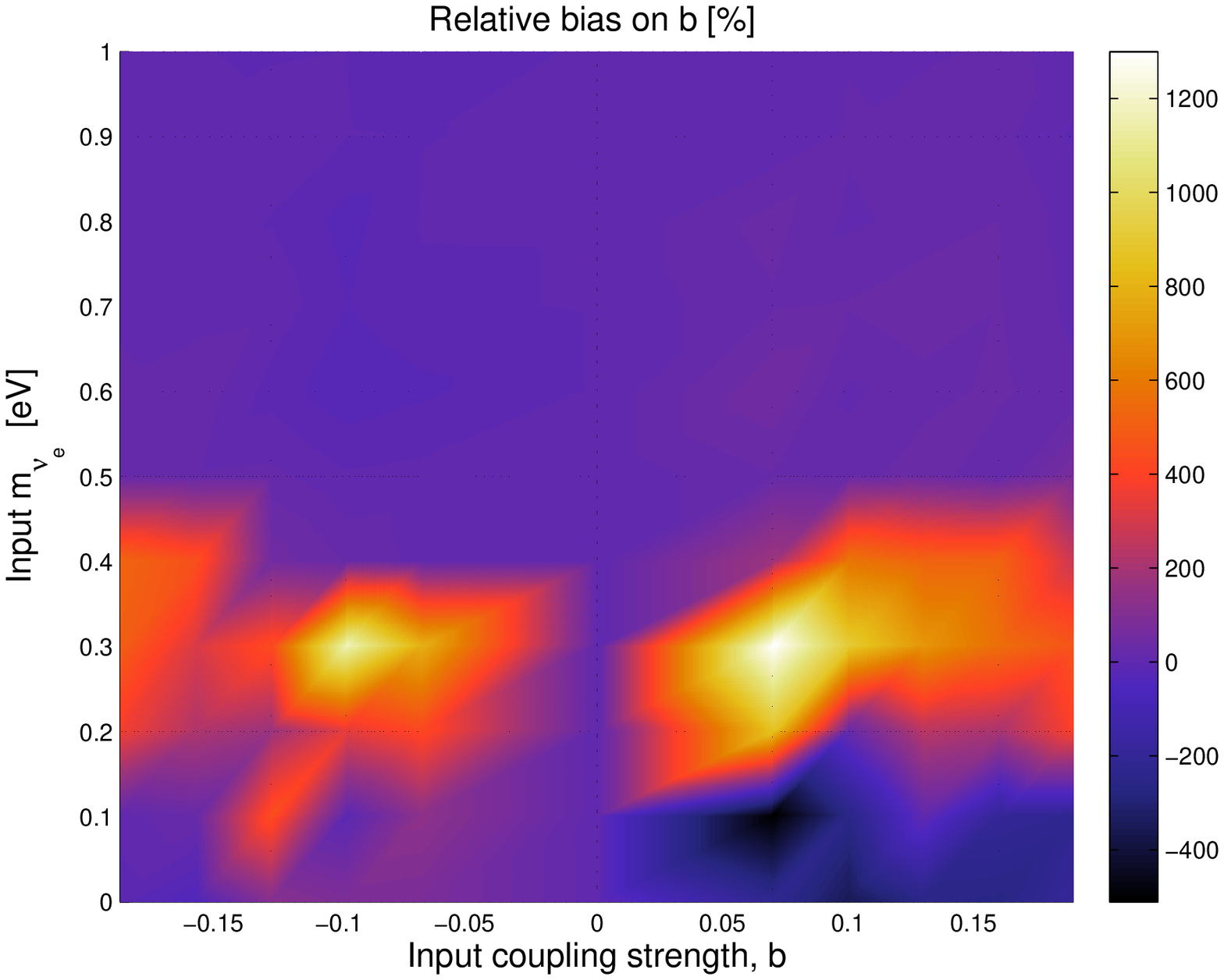}
\hfill
\includegraphics[width=7.0cm]{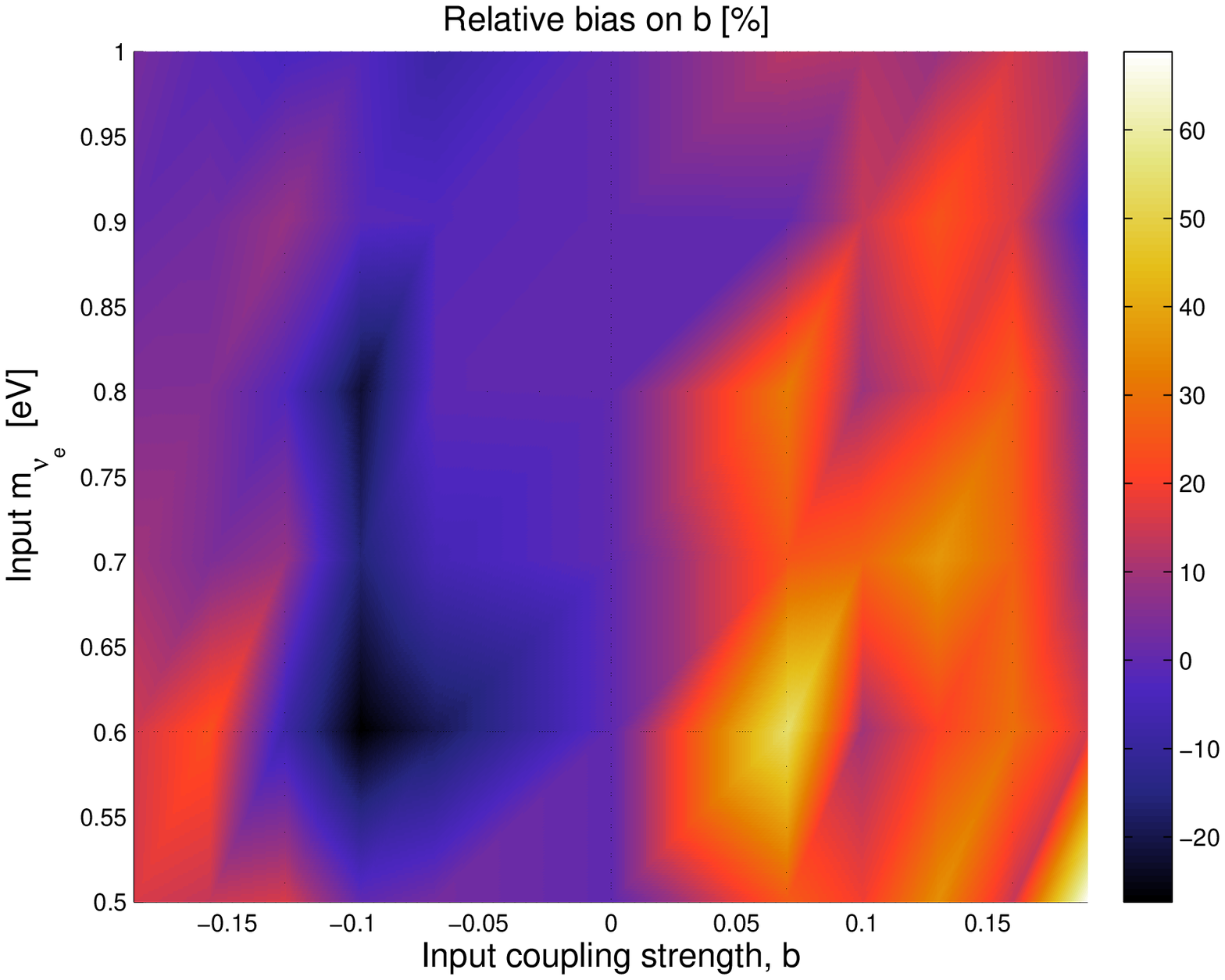}
\caption{The left figure shows the bias on $b$ for the full parameter range of Table \ref{tab:Right} and the right figure is an enlarged version of this plot for $m_{\nu_e}>0.5$ eV. In the left figure we see that we get the output values wrong by more than a factor 10! This is exacerbated at mass values just above the KATRINs sensitivity once again demonstrating how the uncertainty on these two ill-determined parameters is redistributed in the parameter space of the Markov Chain. The right figure shows us that when we look beyond the much larger errorbars around $m_{\nu_e}=0.2$ eV the output values still fluctuate with errors of order $\approx 60 \%$.
 \label{fig:b_r}}
\end{figure}

The results are shown in Figures \ref{fig:m_r} and \ref{fig:b_r}. Note that we included $b$=0.19 to create a better overview of the behavior of $m_{\nu_e}^2$.

\begin{table}
 \centering
  \begin{tabular}[htb!]{ |l | c | c | c | c | c | c | c | c | c | c | r | }
    \hline
    $m_{\nu_e}$ [eV] & 0.0 & 0.1 & 0.2 & 0.3 & 0.4 & 0.5 & 0.6 & 0.7 & 0.8 & 0.9 & 1.0 \\
    \hline
  \end{tabular}
  \vspace*{0.2cm}

  \begin{tabular}[htb!]{ |l | c | c | c | c | c | c | c | c | c | c | r | }
    \hline
        $b$ & -0.19 & -0.16 & -0.13 & -0.10 & -0.07 & 0.0 & 0.07 & 0.10 & 0.13 & 0.16 & 0.19  \\ \hline
   \end{tabular}
  \caption{Input neutrino masses and righthanded coupling strengths used to produce Figures
 \label{tab:Right}}
\end{table}

The results shows us firstly that the output mass values fluctuate rather wildly - and in some cases deviate my as much as $\approx$ 80 \% from the input values as shown in the left panel of Figure \ref{fig:m_r}. And secondly the statistical uncertainty is up to 5 times larger than in the standard case except in regions where $m_{\nu_e}$  < 0.2 eV as expected from the discussion above. Turning to at the output values of the righthanded coupling strength in Figure \ref{fig:b_r} we get appallingly bad results especially in the $m_{\nu_e}$ = 0.2 eV -region. From the left hand picture of Figure \ref{fig:b_r} one might get the impression that the output value of b is returned rather nicely for the larger masses. However as shown in the right panel of Figure \ref{fig:b_r} the relative error is still up to $\approx$ 60\% in some regions. In conclusion we see from these numerical artifacts that it is extremely difficult to get a good determination of {\it{both}} the mass and the coupling strength. At least when using fairly large parameter intervals. Given stronger limits the situation would no doubt change. But judging from our COSMOMC contours the values in some cases will be pressed to the largest allowed parameter values even when the intervals are as broad as here. In other words - tighter parameter values in this case merely amounts to a manual setting of the allowed size of the statistical uncertainties.

\begin{figure}[htb!]
\includegraphics[width=7.0cm]{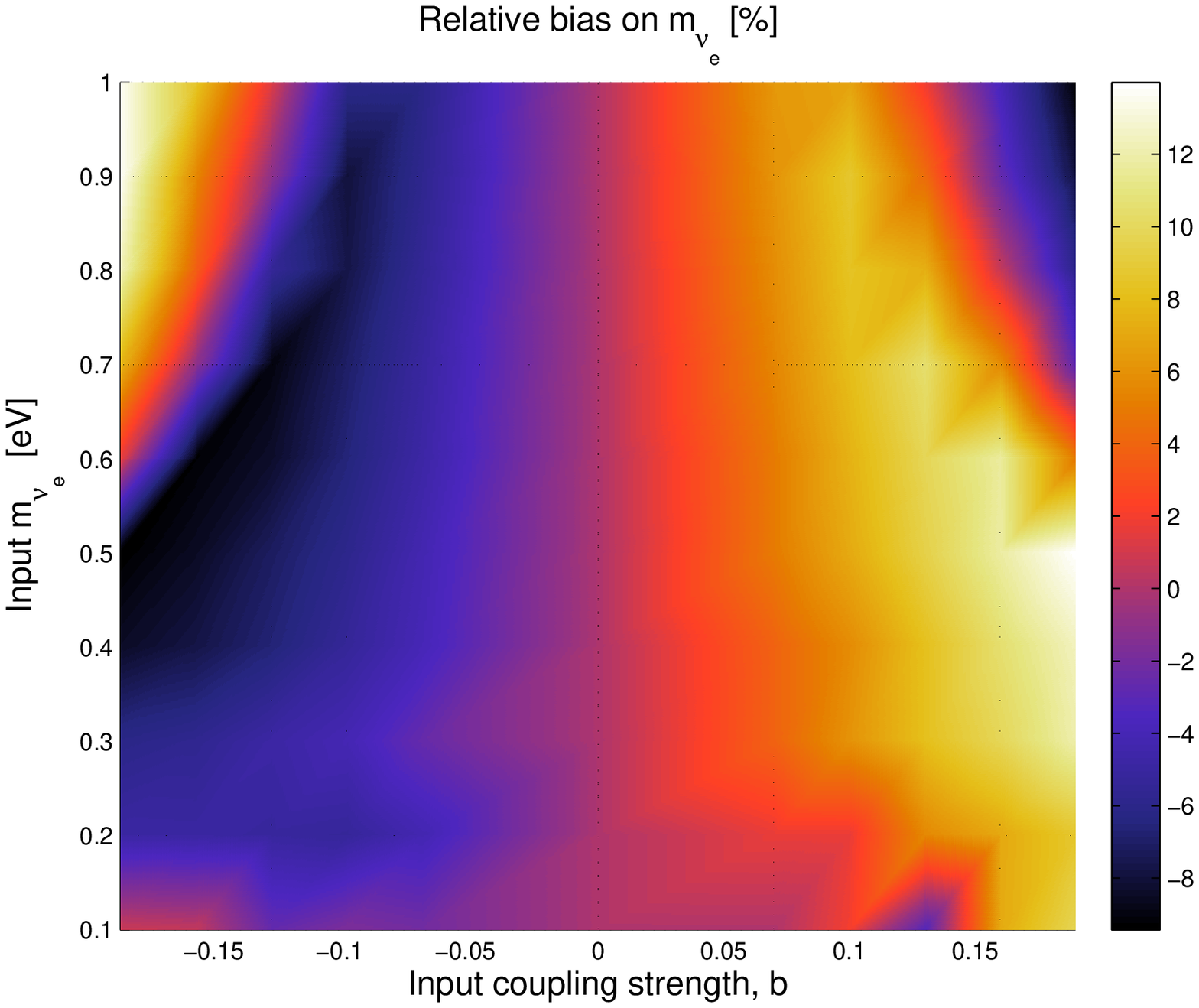}
\hfill
\includegraphics[width=7.0cm]{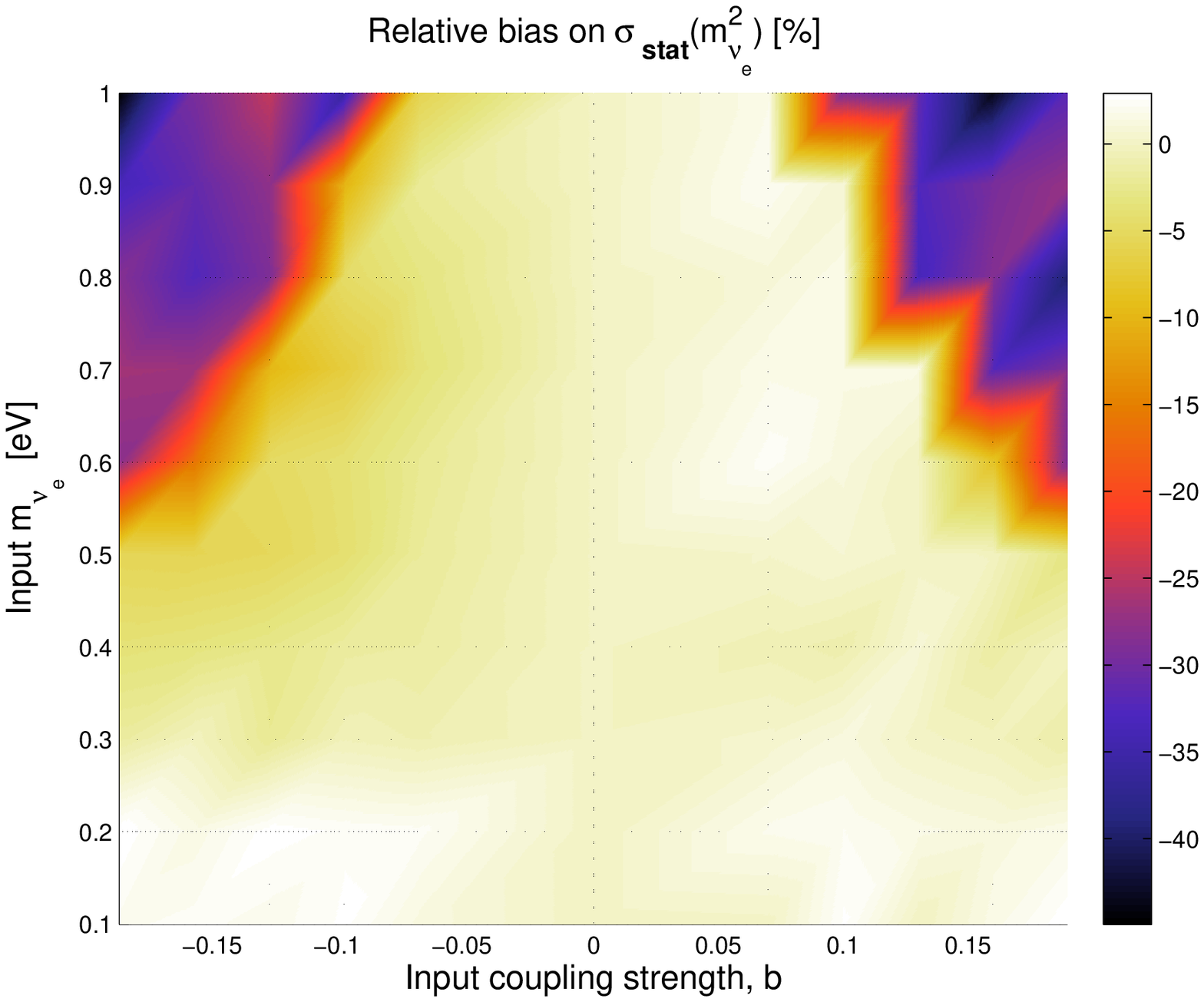}
\caption{This figure shows the same biases as Figure \ref{fig:m_r} but here the analysis has been performed (on the same spectra) without the inclusion of the righthanded coupling strength. Clearly the errors on $m_{\nu_e}$ are much better and for realistic $b$-values certainly within acceptable $\pm10\%$-ranges. However we note that $\sigma_{stat,m_{\nu_e}^2}$ is $\approx 60\%$ better in the high-b, high-m corners of the right-side plot. This coincides with a turnover of the bias on the mass in the left-side plot. Figure \ref{fig:EoR} shows that this behavior takes place because the $\Delta E_0$ parameter is being pushed to the maximally allowed values, which should be avoided. That is the uncertainty on the mass due to the presence of $b$ is migrating into the third correlated parameter - the endpoint of the tritium beta spectrum.
 \label{fig:NoR}}
\end{figure}

Next we perform the analysis on the same spectra without including the righthanded coupling strength to get an idea of the bias imposed on the neutrino mass in the presence of unaccounted-for righthanded currents. We present our results in Figure \ref{fig:NoR}

\begin{figure}[htb!]
\includegraphics[width=7.0cm]{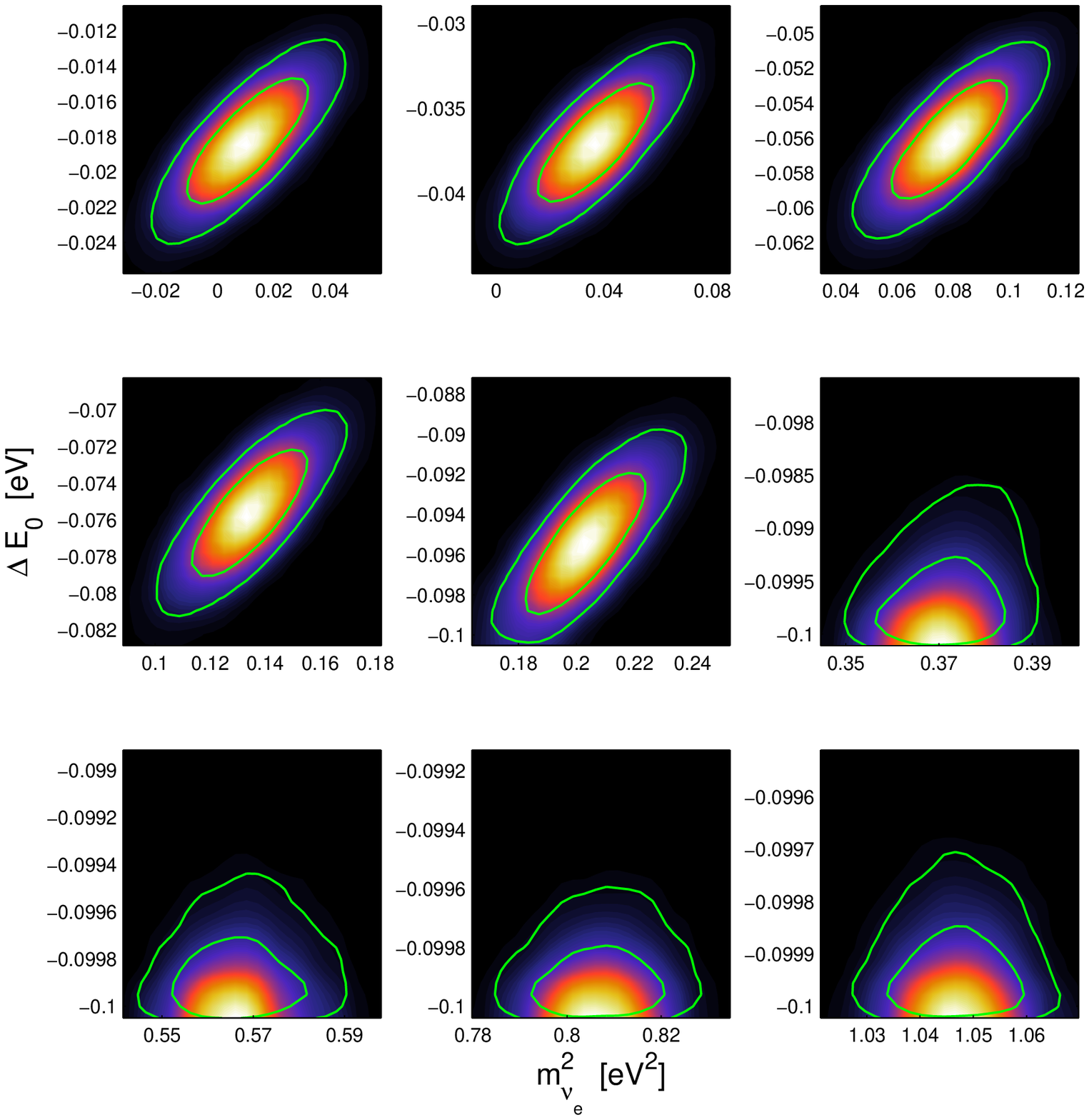}
\hfill
\includegraphics[width=7.0cm]{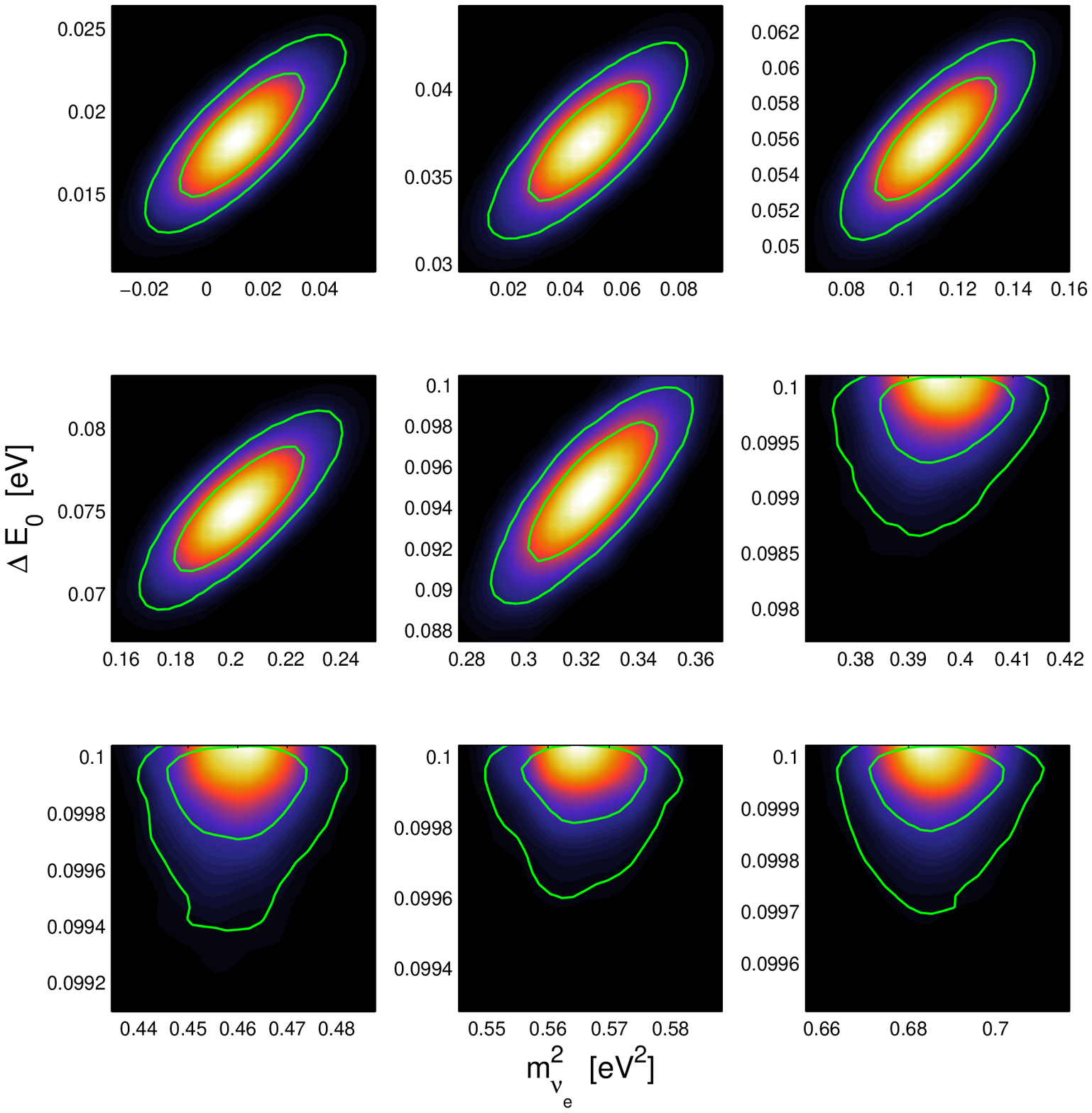}
\caption{The figures shows the 2D likelihood contours of $\Delta E_0$ vs. $m_{\nu_e}^2$ for the mass range $0.1$ eV to $0.9$ eV (again going from the upper left corner to the lower right corner) when the analysis is performed without the inclusion of $b$. The figure on the left used spectra that was produced with $b=-0.19$ while the figure on the right is for $b=0.19$. The expected output for $\Delta E_0$ is zero, but it is clear to see that in this case the $b, m_{\nu_e}^2,\Delta E_0$ -correlation pushes the uncertainty induced in the mass parameter by the physical presence of $b$ into the $\Delta E_0$ -parameter instead. \label{fig:EoR}}
\end{figure}

As it turns out we get much better results when we remove the $b$-dimension from our COSMOMC setup - this time the bias on the mass is no larger than around 12 \%. We notice however that the statistical error drops steeply for high masses and coupling strengths. Inspecting the original COSMOMC likelihood contours we see that this is because $\Delta E_0$ has been pushed to the edge of the input interval as shown in Figure \ref{fig:EoR}. This also explains why the bias flips in the same parameter-range instead of becoming monotonically larger for maximal coupling strengths. The propagation of the uncertainty on $b$ into the $\Delta E_0$-dimension is straightforward from the already discussed correlations between the $b, m_{\nu_e}^2$ and $\Delta E_0$ - parameters. Hopefully the upcoming
much more precise $^3$H--$^3$He mass measurements \cite{blaum:10}  will be helpful in resolving this issue for the KATRIN experiment.

In conclusion the bias induced on the neutrino mass is now within acceptable bounds and agree well with the results found by Bonn {\it{et al}} \cite{Bonn:2007su}.
Finally it should be noted that an experiment such as KATRIN can clearly not be used to put bounds on the size of the righthanded coupling strength at this point. A precise knowledge of the neutrino mass and the tritium beta-spectrum endpoint $E_0$ would be have to be presupposed before measurements of the tritium beta spectrum could be used to determine $b$.

\section{Conclusions}
Our attempt at an analysis of simulated KATRIN data with various additional parameters has shown the following: For the standard case of analysis with regard to one neutrino mass, the MCMC approach is certainly well suited and gives robust results.

The method is very practical when performing analysis for non-standard cases because the COSMOMC output lets us inspect the behavior of the parameters and their relation to one another in a straightforward manner. We have used the method to build a sensitivty-plot for a KATRIN-like experiment, clearly demonstrating the dominating dependence of the sensitivity on the signal countrate.

Further we have learned that for a suitable mass-squared difference an experiment such as KATRIN should be able to detect the existence of other neutrino mass states. And finally we have re-evaluated the influence of couplings to righthanded currents in the tritium beta decay and found that ignoring this would maximally induce an error on the neutrino mass of order 10\%.

In conclusion we find that our bayesian approach to the analysis of the KATRIN experiment is certainly competitive to a frequentist approach and that it has several advantages over it when using an already well-developed framework such as COSMOMC.

\section{Acknowledgements}

We acknowledge the
use of computing resources from the Danish Center for Scientific Computing (DCSC) and
the grant of BMBF under contract 05A08PM1.

\end{document}